\documentclass[12pt, single column, double spaced]{article} % single column, double spaced

\usepackage{ol2}
\usepackage[draft]{hyperref}
\usepackage{amsmath}

\begin{document}

%\twocolumn[%% activate for two-column option

\title{Counting atoms in a deep optical microtrap}

%% For REVTeX it is possible to automate superscript and e-mail callouts with the superscriptaddress option; see REVTeX4 documentation.

\author{Matthew McGovern, Andrew Hilliard, Tzahi Gr\"{u}nzweig and Mikkel F. Andersen$^{*}$}
%\author{M. McGovern, A. J. Hilliard, T. Gr\"{u}nzweig and M. F. Andersen$^{*}$}

\address{Jack Dodd Centre for Quantum Technology, Department of Physics, University of Otago, New Zealand

$^*$Corresponding author: mikkel@physics.otago.ac.nz}

\begin{abstract} We demonstrate a method to count small numbers of atoms held in a deep, microscopic optical dipole trap by collecting fluorescence from atoms exposed to a standing wave of light that is blue detuned from resonance. While scattering photons, the atoms are also cooled by a Sisyphus mechanism that results from the spatial variation in light intensity. The use of a small blue detuning limits the losses due to light assisted collisions, thereby making the method suitable for counting several atoms in a microscopic volume.
\end{abstract}

\ocis{020.1335, 020.3320, 020.7010, 110.0180, 110.2970, 140.7010.}
%]%% activate for two-column option

Atoms in optical microtraps provide a versatile platform for fundamental studies of quantum mechanics at the individual event level and have potential for application in quantum information processing. Recent progress in these fields includes demonstrations of many body quantum states at the single atom level \cite{Bakr2010}, the phase shift of a light beam induced by a single atom \cite{Aljunid2009}, and the realization of a Controlled-NOT quantum gate \cite{Isenhower2010}.

A key challenge in this field is the ability to accurately determine the number of atoms in the optical microtrap. A contemporary technique to achieve this is to collect fluorescence from the atoms when they exposed to polarization gradient cooling (PGC) beams \cite{Miroshnychenko2006a, Bakr2010, Schlosser2001}. However, this technique has several limitations. The cooling beams induce atom loss through light assisted collisions \cite{DePue1999}, often prohibiting the counting of more than one atom \cite{Schlosser2001, Bakr2010, Miroshnychenko2006a}. These beams are also `large' in size and cover all directions,  making it hard to eliminate stray light in detection.

In this Letter, we demonstrate a method to fluorescence image and accurately count small numbers of atoms held in a deep optical microtrap. This is achieved by exposing $^{85}$Rb atoms to a laser beam that is blue detuned from the D1 line at 795~nm and retroreflected to form a standing wave. The use of blue detuned light limits the energy gained in inelastic light-assisted collisions \cite{Bali1994}, such that up to four atoms can be counted. The optical standing wave induces a spatial modulation of the imaging beam intensity, leading to a form of Sisyphus cooling \cite{Dalibard1985}. Unlike other forms of blue detuned cooling \cite{Boiron1996}, this mechanism does not pump the atom(s) into optically dark states, making it ideal for fluorescence imaging. Finally, the use of a dedicated imaging beam several nanometers away from the light used for standard laser cooling at 780~nm, means that a good signal-to-noise ratio may be obtained using standard optical filters to remove stray light from the images.

\begin{figure}[t]
\includegraphics[width=8.3cm]{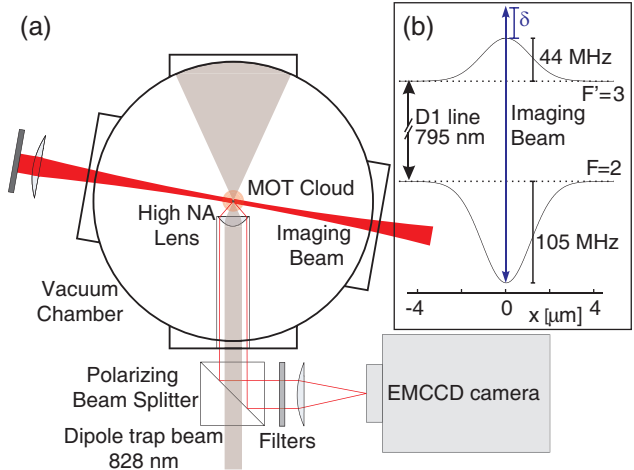}
 \caption{(Color online) (a) Schematic of the experimental setup. (b) Calculated spatially dependent light shifts of
the F = 2 to F'= 3 D1 transition along the tight dimension of the trap. The blue double-headed arrow indicates the frequency of the standing wave imaging light.}
 \label{figure1}
\end{figure}

To prepare small numbers of atoms we load a cloud of atoms from a magneto-optical trap (MOT) into a microscopic dipole trap.
The MOT operates on the F=3 to F'=4 transition of the D2 line in $^{85}$Rb at 780~nm. The atoms are further cooled through 5~ms PGC leaving approximately 50 atoms loaded into the microscopic dipole trap. The dipole trap is formed by focusing a Gaussian beam with wavelength 828~nm  by a high numerical aperture lens (NA = 0.55) to a spot size of $w_0 = 1.8~\mu$m. Figure~\ref{figure1}a is a schematic of the setup. We use 37.0~mW dipole trap power, producing a trap of depth $U_0 = K_{\rm{B}}\times5.03$~mK = $h\times105$~MHz. Figure~\ref{figure1}b shows calculated and experimentally verified \cite{Grunzweig2010} spatially varying light shifts induced by the optical dipole trap. In the following, we quote detunings for an atom at the center of the dipole trap. To reduce the number of trapped atoms we induce light assisted collisions as described in Ref. \cite{Grunzweig2010}.

To image and count small numbers of trapped atoms we induce fluorescence with a retroreflected beam at 795~nm and collect a portion of the light with the high NA lens. The standing wave imaging beam has a  Gaussian intensity profile with waist $w_0 = 92~\mu$m at the position of the atoms. It is linearly polarized and is blue detuned by $\delta=$20~MHz from the D1 F=2 to F'=3 transition (Fig~\ref{figure1}b). The light is applied as a 10~ms pulse, with a rectangular pulse envelope. Atoms that spontaneously decay to the F=3 ground state are pumped back to the F=2 ground state with a 795~nm D1 F=3 to F'=2 beam and the 780~nm PGC beams. The D1 F=3 to F'=2 beam is mode-matched to the imaging beam, tuned to resonance with a power of 10~$\mu$W; this beam can be omitted at the cost of a 44$\%$ decrease in signal. Each PGC beam has a Gaussian intensity profile of waist $\omega_{0}\sim 6$~mm that is apertured by an iris of diameter 8~mm, and for imaging is tuned to resonance on the F=3 to F'=3 D2 line with a power of 1.2~mW. Due to the light shift of the dipole trap, the PGC beams cool atoms on the D2 F=3 to F'=4 transition in the wings of the trap. Approximately half the light collected by the high NA lens is reflected by a polarizing beamsplitter (PBS) located just outside the vacuum chamber. This light passes through optical filters, before a tube lens forms an image on an electron-multiplying charge coupled device (EMCCD) camera. To remove dipole trap and PGC light we use a 795~nm bandpass filter and a notch filter at 830~nm. Additional filtering of room light is effected by encasing the vacuum chamber and associated optics in black-out material. The high NA lens collects 10\% of the fluorescence, and the combined transmission of the PBS, filters and additional optics is 37$\%$. The EMCCD has a measured quantum efficiency of 60$\%$, so that the total photon detection efficiency is 2.3$\%$. We integrate the resulting image and find that we typically collect about 500 photons for one atom, for a 10~ms exposure.

\begin{figure}[t]
\includegraphics[width=8.3cm]{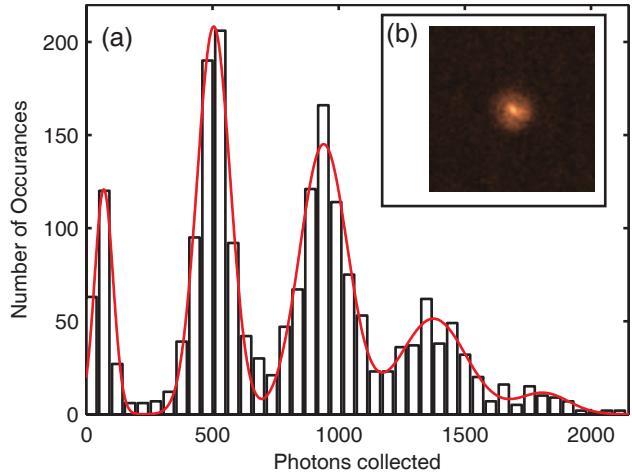}
 \caption{(Color online) (a) Histogram of the integrated fluorescence counts from a small number of atoms in the dipole trap. This experiment was run 2000 times, and the histogram shows resolvable peaks for zero to four atoms from left to right. The red line is a fit of five Gaussians. (b)  2.5~s exposure image of a single atom.}
 \label{figure2}
\end{figure}

The use of blue detuned light for fluorescence imaging allows us to count up to four atoms in the microtrap. Figure \ref{figure2}a shows a histogram of photons collected after the preparation of a small number of atoms. Five peaks corresponding to zero, one, two, three, and four atoms are visible. Broadly, there are two mechanisms at play that allow us to detect more than one atom. The first is `optical shielding': high intensity blue detuned light shields atom pairs from possible inelastic collisions\cite{Bali1994}, reducing the overall inelastic collision rate. Second, if an atom pair undergoes an inelastic light assisted collision, the total energy gained by the pair is limited to the detuning $\delta$, which is less than the trap depth, enforcing the need for multiple collisions to induce atom loss\cite{Hoffmann1996}.
Given that the collision rate scales as $N(N-1)$, where $N$ is the number of atoms, the loss due to light assisted collisions increases sharply for higher numbers of atoms. Figure \ref{figure2}b is a 2.5~s exposure image of a single atom.

\begin{figure}[t]
\centerline{
\includegraphics[width=8.3cm]{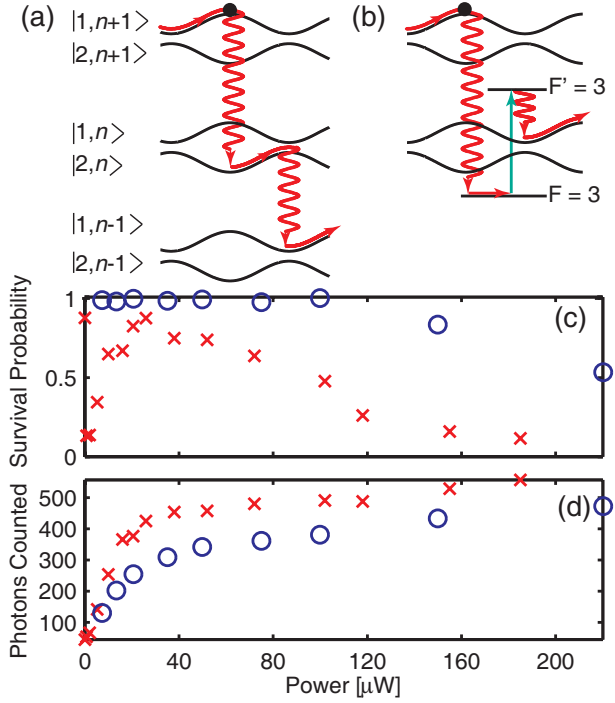}}
 \caption{(Color online) (a) Illustration of the Sisyphus cooling mechanism. Spontaneous transitions between dressed states effect Sisyphus cooling. (b) Atoms that decay to the F = 3 ground state are pumped back into the cooling cycle by the D2 PGC beams (green straight arrow). (c) Survival probability for a single atom as a function of standing wave power for $\delta = 1~$MHz (crosses) and $\delta = 20~$MHz (circles). (d) Integrated fluorescence collected for these two detunings.}
 \label{figure3}
\end{figure}

Inducing fluorescence in atoms with blue detuned laser light may potentially cause Doppler heating as the atoms preferentially pick up the recoil momentum of absorbed co-propagating photons. Heating leads to a loss of atoms from the microtrap, thereby prohibiting imaging and atom counting. To counteract Doppler heating, we employ a variation of the Sisyphus cooling mechanism described in Refs. \cite{Dalibard1985, Aspect1986, Wineland1992}. It relies on two atomic levels, in our case the $5^2S_{1/2}$ F=2 ground state and the D1 $5^2P_{1/2}$ F'=3 excited state coupled by a blue detuned standing wave. The eigenstates of the Schr\"{o}dinger equation for a two-level atom in a near-detuned light field are usually described in terms of the `dressed states' $|1,n\rangle$ and $|2,n\rangle$ of the atom-light system \cite{Cohen-Tannoudji1998a}. $n$ is the total number of photons in the laser field, and in the absence of light, $|1\rangle$ corresponds to the bare atomic ground state, and $|2\rangle$ to the excited state. When an atom moves in the standing wave, the dressed state energy and hence its center-of-mass kinetic energy varies spatially with the local light intensity. The positions of lowest kinetic energy correspond to the highest admixture of excited state in both dressed states, leading to these positions having the highest probability of spontaneous transition between dressed states. This creates a Sisyphus effect, where the atom is predominately moving up potential hills, as illustrated in Fig. \ref{figure3}a. As shown in Ref. \cite{Wineland1992}, the minimum kinetic energy reached by such a cooling mechanism is equal to the depth of the wells created by the standing-wave. Unlike other blue detuned laser cooling mechanisms such as gray molasses, \cite{Boiron1996} where the atoms are pumped into optically dark states, here atoms can scatter many photons and yet remain trapped, thereby enabling us to produce high signal-to-noise images, as in Fig. \ref{figure2}b.

In our case, there is an additional complication to the model considered in Refs. \cite{Dalibard1985,Aspect1986}, since the transition we operate on is not closed. Besides causing the discussed transitions between the dressed states, a spontaneous emission event can also cause an atom to decay into the $5^2S_{1/2}$ F=3 ground state, as indicated in Fig. \ref{figure3}b. The PGC beams will then optically pump the atoms back to the cooling cycle with no preference to position along the standing wave. Therefore, this does not qualitatively change the picture given above.

The question of whether Sisyphus cooling or Doppler heating dominates depends on the intensity of the light used to form the standing wave. The energy increase per photon scattered in Doppler heating is independent of the light intensity. On the other hand, the Sisyphus energy loss per photon depends on the magnitude of the potential hills the atom climbs. Sisyphus cooling therefore dominates at high intensities and Doppler heating at low intensities.

To investigate luminosity and `survival probability' of the atom as a function of standing wave parameters we ran the following set of experiments. The D1 F=3 to F'=2 beam was omitted here to isolate the role of the blue detuned standing wave. In 200 repetitions of a control experiment, we prepared single atoms, and took two 10~ms exposure images, 30~ms apart with $\delta=20~$MHz and a power of $30~\mu$W for the standing wave. This yielded for these imaging parameters the probability of detecting an atom in the final image ($F$), conditioned on it being detected in the initial image ($I$): $P_{c}(F|I) = 0.97$. To test the detuning and power dependence of imaging, we took three 10~ms exposure images of single atoms 10~ms apart, where the first and last images were taken under the same conditions as in the control experiment, and the standing wave parameters for the second image were varied. We define the survival probability as $P_{x}(F|I) / P_{c}(F|I)$, where $P_{x}(F|I)$ is the probability of having an atom in the final image conditioned on having it present in the initial image.
To measure the survival probability, we ran 200 repetitions at each detuning and power and show the result in Fig. \ref{figure3}c.
For a detuning of $\delta=1$~MHz, we see a low survival probability at low powers which we attribute to Doppler heating dominating over Sisyphus cooling. For larger powers, the survival probability increases, as Sisyphus cooling becomes dominant. For powers above 75~$\mu$W the depth of the standing wave wells become comparable to the depth of the dipole trap. As the atom's typical external energy is set by the depth of the standing wave wells, this leads to atom loss and manifests itself in the decreasing survival probability observed. A similar trend is observed for $\delta=20~$MHz but the trapped atom has a high survival probability for a larger range of imaging beam powers. Figure \ref{figure3}d shows the integrated fluorescence counts from the second image from the runs where an atom remained in all three images. The fluorescence counts for detuning $\delta=1~$MHz are higher than the counts for detuning $\delta=20$~MHz because it is closer to resonance. Therefore, one must compromise between high fluorescence and atom loss. $\delta=20$~MHz and a power of 50~$\mu$W comprise a useful parameter set with a survival probability of 99$\%$ in combination with a relatively high fluorescence count.

In conclusion, we have demonstrated a technique to count small numbers of atoms in an optical microtrap. By using a dedicated blue detuned standing wave to induce fluorescence, we reduce the energy available in light assisted collisions between atoms, making it possible to count up to four atoms in the trap. We employ a Sisyphus cooling method to cool the atoms while they fluoresce, thereby counteracting Doppler heating, and find parameters where it is effective. Our work may open new avenues in few body physics with neutral atoms and the study of molecule formation.

This work is supported by NZ-FRST Contract No. NERF-UOOX0703 and UORG.

% [use \verb+\pagebreak+ to balance final column]

\bibliographystyle{unsrt}
%\bibliography{SingleAtom}

\end{document}